\documentclass[11pt]{article}
\usepackage[margin=2cm]{geometry}
\usepackage{authblk}
\usepackage{amsmath}
\usepackage[utf8]{inputenc}
\usepackage[T1]{fontenc}
\usepackage{xcolor}
\usepackage{paracol}
\usepackage{graphicx}
\usepackage{placeins} % To force figures before next paragraph
\usepackage{cleveref}

\newcommand{\be}{\begin{equation}}
\newcommand{\ee}{\end{equation}}

\definecolor{ao}{rgb}{0.0, 0.0, 1.0}
\definecolor{blue(pigment)}{rgb}{0.2, 0.2, 0.6}

\title{Spherically symmetric density and potential of a hydrogen molecule}
\date{}
\author[1]{K. Kokko} 
\author[2]{\'A. Nagy}
\author[1]{J. Huhtala}
\author[3]{T. Bj\"orkman}
\author[4,5,6]{L. Vitos}
\affil[1]{Department of Physics and Astronomy, University of Turku, FI-20014 University of Turku, Finland}
\affil[2]{Department of Theoretical Physics, University of Debrecen, H-4002 Debrecen, Hungary }
\affil[3]{Faculty of Science and Engineering, \AA bo Akademi University, FI-20500 Turku, Finland}
\affil[4]{Applied Materials Physics, Department of Materials Science and Engineering, Royal Institute of Technology}
\affil[5]{Department of Physics and Astronomy, Division of Materials Theory, Uppsala University, Box 516, SE-75121 Uppsala, Sweden}
\affil[6]{Research Institute for Solid State Physics and Optics, Wigner Research Center for Physics, P.O. Box 49, H-1525 Budapest, Hungary)}

\begin{document}
\maketitle

\begin{abstract}
Using a hydrogen molecule as a test system we demonstrate how to compute the effective potential according to the formalism of the new density functional theory (DFT), in which the basic variable is the set of spherically averaged densities instead of the total density,  used in the traditional DFT. The effective potential together the external potential, nuclear Coulomb potential, can be substituted in the Schr\"odinger like differential equation to obtain the spherically averaged electron density of the system. In the new method instead of one three-dimensional low symmetry equation one has to solve as many spherically symmetric equations as there are atoms in the system.
\end{abstract}

More than fifty years have passed since the fundamental theorems behind one of the most successful quantum theories of the electron gas was put forward \cite{Hohenberg_1964}, \cite{Kohn_1965}. This formulation of the density functional theory (DFT) of the 60s has enabled a large, still growing, number of significant theoretical and computational studies of materials properties. The new formulation of DFT, first published in 2018 by Theophilou \cite{Theophilou_2018}, states that besides the total electron density there exists also a set of spherically symmetric densities that determines the external potential of the system. A different approach for using spherically symmetric densities as basic variables was introduced by Nagy \cite{Nagy_2018}.  The spherically symmetric  approach could lead to dramatic improvements of the computational methods.

In her recent article \cite{Nagy_2020} \'A.\ Nagy formulates a Schr\"odinger-like differential equation for the spherically symmetric densities of many atom systems. Based on the results presented in Ref.\ \cite{Nagy_2020} we study here how the spherically symmetric procedure can be implemented in numerical formalism. For the test case we choose the hydrogen molecule. Hartree units are used exclusively.

The electronic wave function of the H$_2$ molecule we represent by a $2\times2$ determinant. Based on that choice we demonstrate how to construct the Schr\"odinger-like equation for numerical calculations of the electronic density of H$_2$.

For this purpose the key equations in Ref.\ \cite{Nagy_2020} are
\be
\hat{h}_\textrm{eff}^\alpha (r_\alpha)\sigma_{\alpha}^{1/2}(r_{\alpha}) = \mu\sigma_{\alpha}^{1/2}(r_{\alpha})
\ee
where
\be
\hat{h}_\textrm{eff}^\alpha (r_\alpha) = -\frac{1}{2}\frac{\mathrm{d}^2}{\mathrm{d}r_\alpha^2}
+ \nu_\alpha(r_\alpha) + \nu_{\rm eff}^\alpha(r_\alpha) 
\label{eq:hamilton}
\ee
and 
\be
\nu_\alpha( r_\alpha) = -\frac{Z_\alpha}{r_\alpha}, \,\, r_\alpha = |\mathbf{r} - \mathbf{R}_\alpha| 
\ee
\begin{equation}
\sigma_{\alpha}(r_{\alpha}) = 4\pi {r_{\alpha}}^2\bar{\rho}_{\alpha}(r_{\alpha}) 
\end{equation}
where
\be
\bar{\rho}_{\alpha}(r_{\alpha}) = \frac{1}{4\pi}\int_{{\Omega}_{\alpha}}\rho(\mathbf{r}_{\alpha}) \mathrm{d}\Omega_{\alpha}
\ee
\be
\rho(\mathbf{r_\alpha}) = N\int |\Psi(x_1, ..., x_{N-1},\mathbf{r}_\alpha, s_N)|^2 \mathrm{d}x_1 ... \mathrm{d}x_{N-1}\mathrm{d}s_N
\ee
where $N$ is the number of electrons, $\alpha$ refers to a nucleus ($\alpha \in \{1,... M\}$), $Z_{\alpha}$ is th charge of the nucleus $\alpha$, $\mathbf{r}_\alpha$ is the position of an electron with respect to the nucleus $\alpha$, $s$ is the spin variable and $\mu$ is the chemical potential, the negative of the ionization energy $I$. $\Psi(x_1, ..., x_{N-1},\mathbf{r}_\alpha, s_N)$ is the total electronic wave function of the many atom system. There are $M$ equations for the spherically symmetric densities (one for each nucleus).

The effective potential $\nu_{\rm eff}^\alpha(r_\alpha)$ can be broken down into several parts (in the following formulas we use $N=2$ for H$_2$):
\begin{eqnarray}
\hspace{-8ex}\nu_{\rm eff1}^\alpha(r_\alpha) &=& \underbrace{\int \Phi^*(x_1,\hat{r}_\alpha)[\hat{H}_{1}-E_{1}^0] 
\Phi(x_1,\hat{r}_{\alpha})\mathrm{d}x_1\mathrm{d}\hat{r}_{\alpha} \mathrm{d}s_2}_{\nu_{\rm eff1,1}^\alpha(r_\alpha)} 
+ \underbrace{\int\frac{1}{|\mathbf{r}_1-\mathbf{r}_\alpha|}D(\mathbf{r}_1,\mathbf{r}_\alpha)\mathrm{d}\mathbf{r}_1\mathrm{d}\hat{r}_\alpha}_{\nu_{\rm eff1,2}^\alpha(r_\alpha)} \nonumber \\
&+& \underbrace{\frac{1}{2}\int |\nabla_{\mathbf{r}_\alpha}\Phi(x_1,\hat{r}_{\alpha})|^2 \mathrm{d}x_1\mathrm{d}\hat{r}_\alpha \mathrm{d}s_2}_{\nu_{\rm eff1,3}^\alpha(r_\alpha)}
\end{eqnarray}
\begin{eqnarray}
\hspace{-6ex}\nu_{\rm eff2}^\alpha(r_\alpha) &=& \underbrace{-\frac{1}{2r_\alpha^2}\int \Phi^*(x_1,\hat{r}_\alpha)\nabla_{\theta_\alpha \phi_\alpha}^2 
 \Phi(x_1,\hat{r}_\alpha)\mathrm{d}x_1\mathrm{d}\hat{r}_\alpha \mathrm{d}s_2}_{\nu_{\rm eff2,1}^\alpha(r_\alpha)} \nonumber \\
 &+& \underbrace{\int \nu_1(|\mathbf{r}_\alpha - \mathbf{R}_{21}|)
|\Phi(x_1,\hat{r}_\alpha)|^2 
\mathrm{d}x_1\mathrm{d}\hat{r}_\alpha \mathrm{d}s_2}_{\nu_{\rm eff2,2}^\alpha(r_\alpha)}
\end{eqnarray}
where
\be
\hspace{-8ex}D(\mathbf{r}_1,\mathbf{r}_{\alpha}) = \int|\Phi(x_1,\hat{r}_{\alpha},s_2)|^2 \mathrm{d}s_1 \mathrm{d}s_2
\ee
\be
 \nu_1(|\mathbf{r}_\alpha - \mathbf{R}_{21}|)) = \frac{-1}{|\mathbf{r}_\alpha - \mathbf{R}_{21}|}
\ee
and  $\Phi$ is defined as
\be
\Psi(x_1, x_2) = (2\pi)^{1/2}[\bar\rho_{\alpha}(r_{\alpha})]^{1/2}\Phi(x_1, \hat{r}_{\alpha},s_2).
\label{factors}
\ee

The one-electron orbital is a function of the space coordinates ($x_{\mathrm{i}}$, $y_{\mathrm{i}}$, $z_{\mathrm{i}}$) and the spin function $\sigma_i$ has one of two values ($a$ or $b$). The electronic wave function of H$_2$ is then written as a Slater determinant
\be
\Psi = \frac{1}{\sqrt{2}}
\begin{vmatrix} 
(\psi_{\gamma_1}\sigma_1)^1 & (\psi_{\gamma_1}\sigma_1)^2 \\ (\psi_{\gamma_2}\sigma_2)^1 & (\psi_{\gamma_2}\sigma_2)^2
\end{vmatrix}.
\ee
Orbitals $\psi_\gamma$, are linear combinations of functions $\phi_i$ and the superscript shows the electron occupying the spin orbital in question. 
\be
\psi_\gamma = \sum_i C_{\gamma i}\phi_i.
\ee

 We use a minimum basis set composed of two 1s atomic orbitals
\begin{eqnarray}
\phi_1 =  \frac{1}{\sqrt{\pi}}\exp(-r_\mathrm{A}) \nonumber \\
\phi_2 =  \frac{1}{\sqrt{\pi}}\exp(-r_\mathrm{B}),
\end{eqnarray}
where $r_\mathrm{A}$ and $r_\mathrm{B}$ are the distances of the electrons from the respective nuclei ($A$ and $B$). The ground state is constructed using a molecular orbital
\be
\psi_1 = N_1(\phi_1 + \phi_2),
\label{eq:Psi_1}
\ee
\begin{figure}[ht]
\includegraphics[width=0.4\textwidth, angle =- 90]{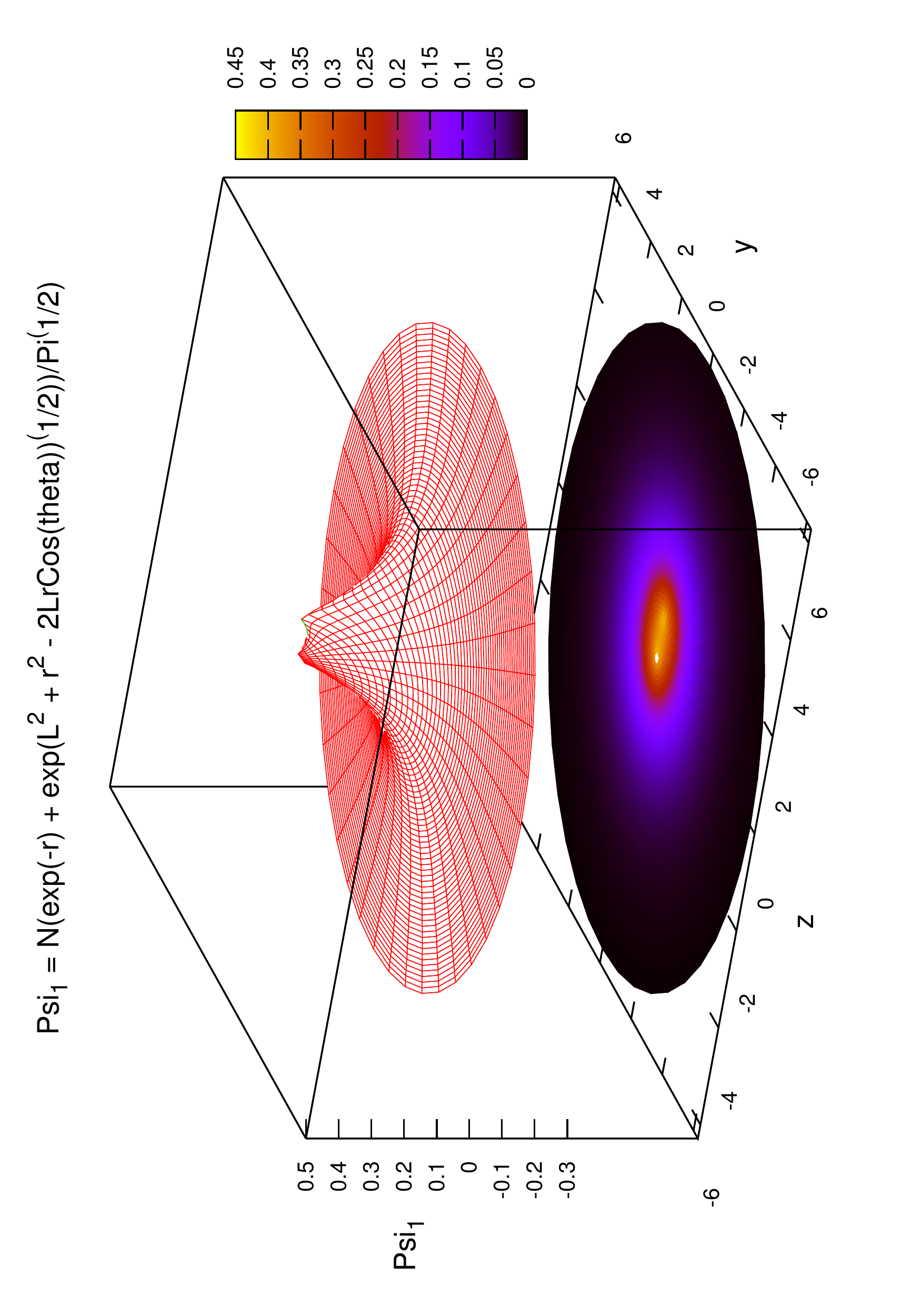}
\caption{\label{fig:Psi_1}Molecular orbital $\psi_1$, Eq. (\protect\ref{eq:Psi_1}) plotted in ($y$,$z$) plane. Nuclei are in the $z$ axis, one at the origin and the other one at $z = 0.84$ ($y$ and $z$ are given in Bohr radii).}
\end{figure}
where normalization factor $N_1$ is
\be
N_1 = \left(\int\psi_1^2\mathrm{d}r\right)^{-1/2}. 
\ee

\FloatBarrier
To exploit the mathematical equation (\ref{factors}) we factorize the wave function (\ref{eq:Psi_1}) into radial and angular parts by minimizing the difference $|N_1(\phi_1 + \phi_2) -\bar\rho_{\alpha} (r_{\alpha})^{1/2}A(\theta_{\alpha})|$. The optimized (unnormalized) $A(\theta)$ is
\be
A(\theta_{\alpha}) =  p(q + s\cos[((\arccos[\cos[\theta_{\alpha}]] - \pi)/\pi)^7+1]),
\ee
where $p$, $q$, $s$ are 0.8, 1.7, and 2.0,respectively,  giving 
\be
\psi_1 = \bar\rho_{\alpha} (r_{\alpha})^{1/2}A(\theta_{\alpha})\left[  \int A(\theta_{\alpha})^2 \mathrm{d}\Omega_{\alpha}\right]^{-1}.
\ee

\begin{figure}
\includegraphics[width=0.5\textwidth, angle =- 90]{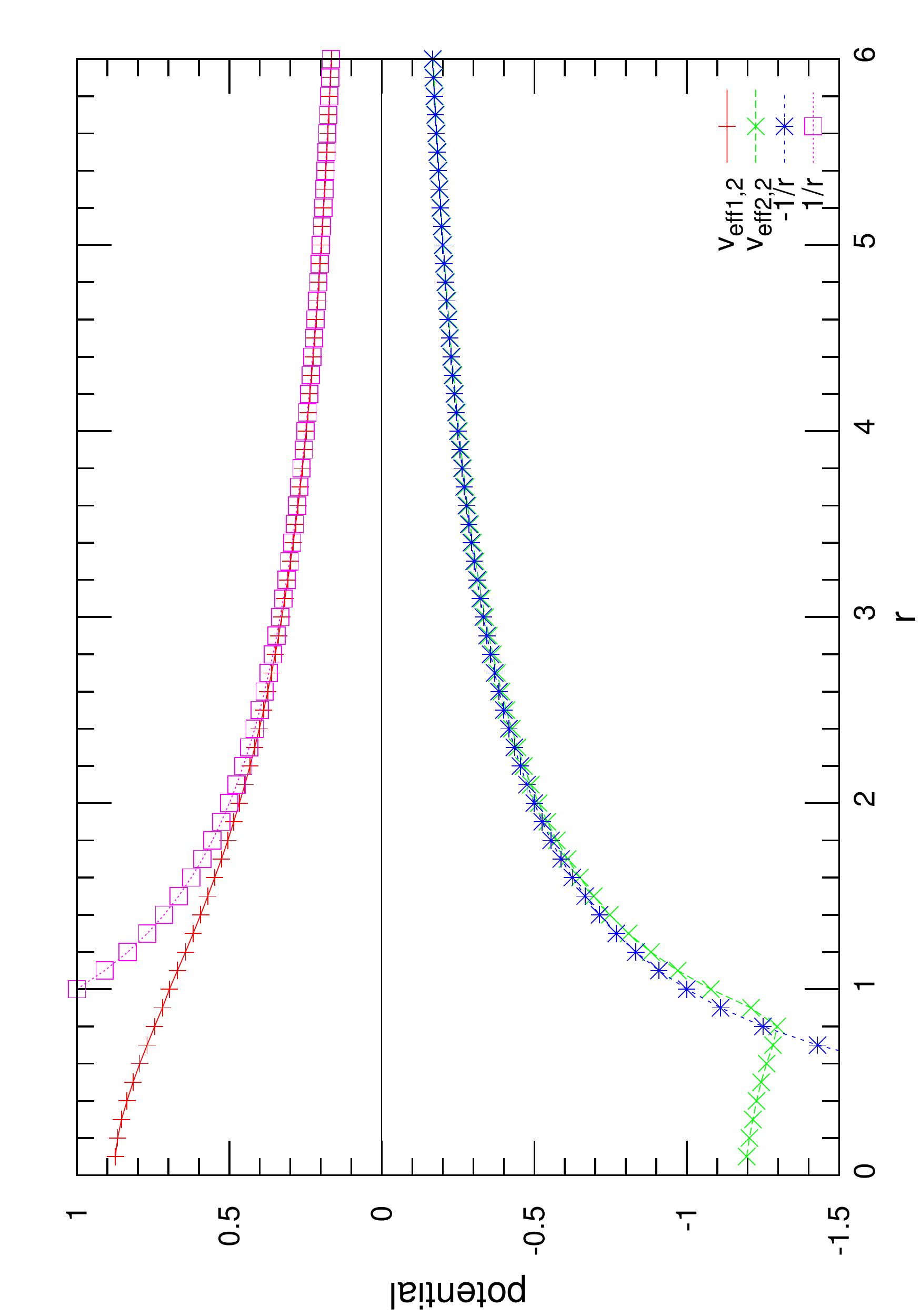}
\caption{\label{fig:v12v22_20200911}Potentials $v_{\mathrm{eff1,2}}$ Eq.\ (\protect\ref{eq:v12}) and $v_{\mathrm{eff2,2}}$ Eq.\ (\protect\ref{eq:v22})  ($-1/r$ and $1/r$ are shown for comparison). $r$ is given in Bohr radii and potential in Hartrees.}
\end{figure}

\begin{figure}
\includegraphics[width=0.5\textwidth, angle =- 90]{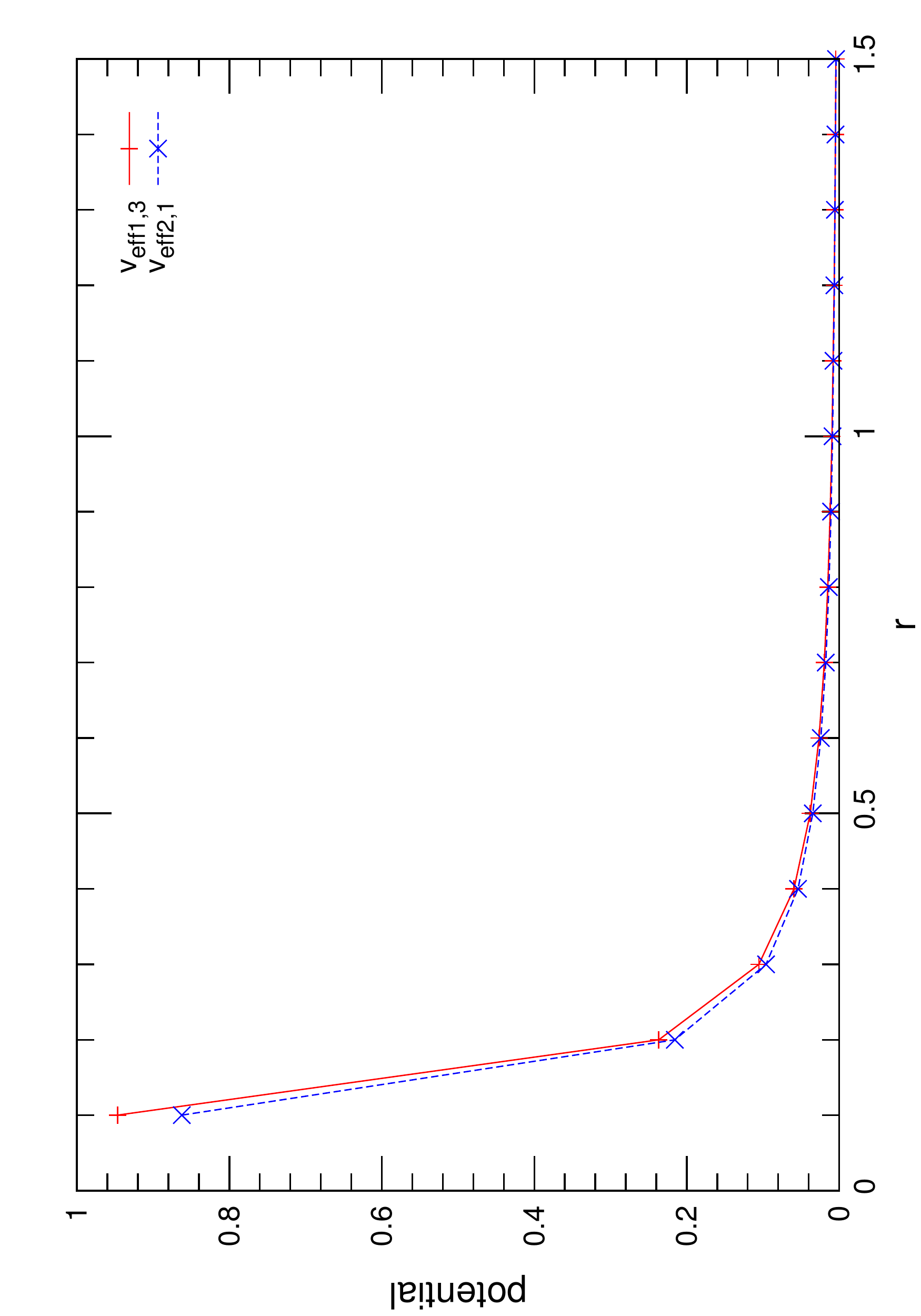}
\caption{\label{fig:v13v21_20200912}Potentials $\nu_\mathrm{eff1,3}^{\alpha}(r_{\alpha})$ Eq. (\protect\ref{eq:v13}) and $\nu_\mathrm{eff2,1}^\alpha(r_\alpha)$ Eq. (\protect\ref{eq:v21}). $r$ is given in Bohr radii and potential in Hartrees.}
\end{figure}

The above procedure leads to the following expressions for the potential terms:
\be
\nu_\mathrm{eff,1,1}^{\alpha}(r_{\alpha}) \approx E_1-E_{N-1}^0 + N_1^2 <\phi_2(\mathbf{r_{1}})| v_{\alpha} - v_{\beta} | \phi_2(\mathbf{r_{1}}) > ,
\label{eq:v11}
\ee
where $E_1$ and $E_{N-1}^0$ are the ground state energy of the hydrogen atom and the ground state energy of the singly ionized hydrogen molecule, respectively
\begin{eqnarray}
& &\nu_\mathrm{eff,1,2}^{\alpha}(r_{\alpha})\nonumber \approx \\ &\hspace{-15 ex} &\hspace{-10 ex}\Bigg{[} \hspace{-1 ex} \int \hspace{-1 ex} \Bigg{\{}\int\hspace{-2 ex}\frac{1}{\sqrt{[r_{\alpha}\cos(\theta_{\alpha})-r_1\cos(\theta_1)]^2 + [r_{\alpha}\sin(\theta_{\alpha})\cos(\phi_{\alpha})-r_1\sin(\theta_1)\cos(\phi_1)]^2  +[r_{\alpha}\sin(\theta_{\alpha})\sin(\phi_{\alpha})-r_1\sin(\theta_1)\sin(\phi_1)]^2 }} 
\nonumber \\
& & \mathrm{d}\phi_1\mathrm{d}\phi_{\alpha}\Bigg{\}} \psi(r_1,\theta_1)^2r_1^2\sin(\theta_1) A(\theta_{\alpha})^2 \sin(\theta_{\alpha})  \mathrm{d}r_1\mathrm{d}\theta_1\mathrm{d}\theta_{\alpha}\Bigg{]}  \left[  \int A(\theta_{\alpha})^2 \sin(\theta_{\alpha})\mathrm{d}\theta_{\alpha}\right]^{-1}
\label{eq:v12}
\end{eqnarray}
\be
\nu_\mathrm{eff,1,3}^\alpha(r_\alpha) \approx \frac{1}{2}\frac{1}{r_\alpha^2}\int \left[\frac{\partial A(\theta_\alpha)}{\partial \theta_\alpha}\right] ^2\mathrm{d}\Omega_\alpha  \left[ \int A(\theta_\alpha)^2 \mathrm{d}\Omega_\alpha\right]^{-1} 
\label{eq:v13}
\ee
\be
\nu_\mathrm{eff2,1}^\alpha(r_\alpha) \approx -\frac{1}{2r_\alpha^2}\int  \left[ \int A(\theta_\alpha)^2 \mathrm{d}\Omega_\alpha\right]^{-1}
 \int A(\theta_\alpha)\nabla_{\theta_\alpha \phi_\alpha}^2A(\theta_\alpha)\, \mathrm{d}\Omega_\alpha
 \label{eq:v21}
\ee
\be
\nu_\mathrm{eff2,2}^\alpha(r_\alpha) \approx \left[ \int A(\theta_\alpha)^2 \mathrm{d}\Omega_\alpha\right]^{-1}   \int  \frac{-A(\theta_\alpha)^2  }{ \sqrt{(L - r_\alpha\cos{\theta_\alpha})^2 + (r_\alpha\sin{\theta_\alpha})^2}}\mathrm{d}\Omega_\alpha
\label{eq:v22}
\ee
The potential terms $\nu_\mathrm{eff1,2}^{\alpha}(r_{\alpha})$ and $\nu_\mathrm{eff2,2}^\alpha(r_\alpha)$ are shown in Fig.\ \ref{fig:v12v22_20200911}, the electrostatic potentials $\frac{-1}{r}$  and $\frac{1}{r}$ are shown for comparison. Fig. \ref{fig:v13v21_20200912} shows the potentials $\nu_\mathrm{eff1,3}^{\alpha}(r_{\alpha})$ and $\nu_\mathrm{eff2,1}^\alpha(r_\alpha)$.

\begin{figure}
\includegraphics[width=0.5\textwidth, angle = -90]{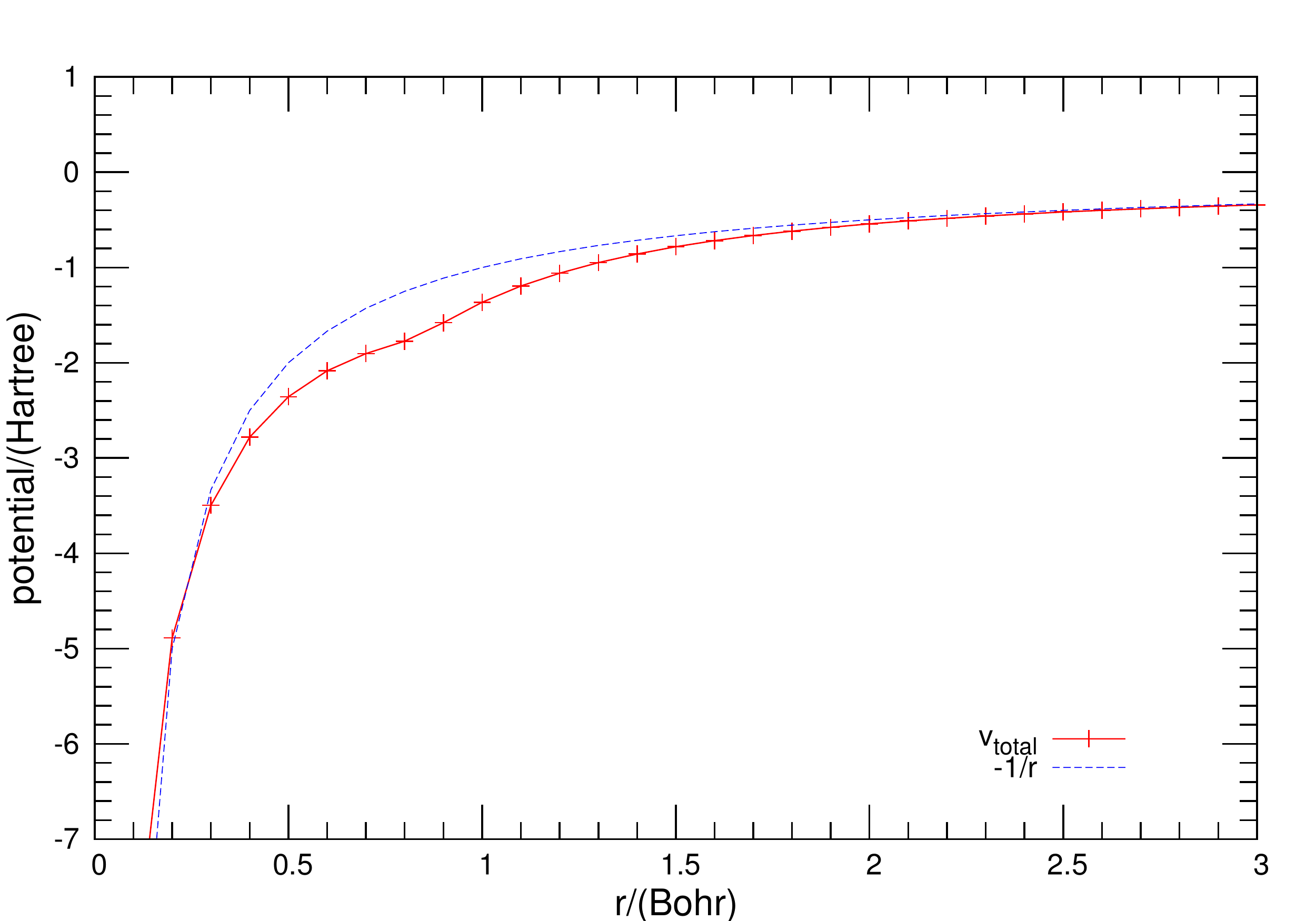}
\caption{\label{fig:v_total_small_20200915} The red curve shows the sum of potentials ($v_{\mathrm{total}}$) from equations  (\protect\ref{eq:v12}), (\protect\ref{eq:v13}), (\protect\ref{eq:v21}), (\protect\ref{eq:v22}), and $-1/r$. For comparison also the potential of one nucleus $-1/r$ is shown by separate curve (blue).}
\end{figure}

Fig. \ref{fig:v_total_small_20200915} shows the sum of the potentials obtained from Eqs. (\ref{eq:v12}), (\ref{eq:v13}), (\ref{eq:v21}), (\ref{eq:v22}), and $-1/r$. The 'other' nucleus can be seen to lower the potential in the region $r \in (0.3, 2)$ and to increase the potential when $r < 0.2$. The position of the 'other' nucleus is at $r = 0.84$ The effect in the region $r \in (0.3, 2)$ is mainly due to potentials  $\nu_\mathrm{eff1,2}^{\alpha}$ and $\nu_\mathrm{eff2,2}$ (Fig.\ \ref{fig:v12v22_20200911}) and the effect in the region $r < 0.2$ is mainly due to the potentials $\nu_\mathrm{eff1,3}^{\alpha}$ and $\nu_\mathrm{eff2,1}^\alpha$ (Fig.\ \ref{fig:v13v21_20200912} ).

\begin{thebibliography}{001}
\bibitem{Hohenberg_1964}Hohenberg, P. and Kohn, W., Physical Review. 136 (3B): B864?B871
(1964).
\bibitem{Kohn_1965}Kohn, W. and Sham, L. J., Phys. Rev. 140, A1133 (1965).
\bibitem{Theophilou_2018}Theophilou, J., Chem. Phys. 149, 074104 (2018).
\bibitem{Nagy_2018} Nagy, \'A., J. Chem. Phys. 149, 204112 (2018).
\bibitem{Nagy_2020}Nagy, \'A., J. Phys. Chem. A 2020, 124, 148 -- 151.
\end{thebibliography}
 \end{document}